\documentclass[fleqn,usenatbib,useAMS]{mnras}

\usepackage{graphicx}	
\usepackage{amsmath}	
\usepackage{amssymb}	
\usepackage{multicol}        
\usepackage{bm}		
\usepackage{pdflscape}	
\usepackage{csvsimple}

\usepackage[T1]{fontenc}
\usepackage{ae,aecompl}

\usepackage{newtxtext,newtxmath}

\title[Pulsars with the UWL]
{Two years of pulsar observations with the Ultra-Wideband Receiver on the Parkes radio telescope}
\author[Johnston et al.]
{
Simon Johnston$^{1}$\thanks{E-mail: simon.johnston@csiro.au},
C. Sobey$^2$,
S. Dai$^{1}$,
M. Keith$^{3}$,
M. Kerr$^{4}$,
R. N. Manchester$^{1}$,
L. S. Oswald$^{5}$, \newauthor
A. Parthasarathy$^{6}$,
R. M. Shannon$^{7,8}$,
P. Weltevrede$^{3}$
\\
$^{1}$CSIRO Astronomy and Space Science, Australia Telescope National Facility, PO~Box~76, Epping NSW~1710, Australia\\
$^{2}$CSIRO Astronomy and Space Science, PO Box 1130, Bentley, WA 6102, Australia\\
$^{3}$Jodrell Bank Centre for Astrophysics, The University of Manchester, Alan Turing Building, Manchester M13 9PL, UK\\
$^{4}$Space Science Division, Naval Research Laboratory, Washington, DC 20375, USA\\
$^{5}$Department of Astrophysics, University of Oxford, Denys Wilkinson Building, Keble Road, Oxford OX1 3RH, UK\\
$^{6}$Max-Planck-Institut f{\"u}r Radioastronomie, Auf dem H{\"u}gel 69, D-53121 Bonn, Germany\\
$^{7}$Centre for Astrophysics and Supercomputing, Swinburne University of Technology, PO Box 218, Hawthorn, VIC 3122, Australia\\
$^{8}$OzGrav: The ARC Centre of Excellence for Gravitational-wave Discovery, Hawthorn VIC 3122, Australia
}
\date{Last updated; in original form}

\pubyear{2020}

\begin{document}
\label{firstpage}
\pagerange{\pageref{firstpage}--\pageref{lastpage}}
\maketitle

\begin{abstract}
The major programme for observing young, non-recycled pulsars with the Parkes telescope has transitioned from a narrow-band system to an ultra-wideband system capable of observing between 704 and 4032~MHz. We report here on the initial two years of observations with this receiver. Results include dispersion measure (DM) and Faraday rotation measure (RM) variability with time, determined with higher precision than hitherto, flux density measurements and the discovery of several nulling and mode changing pulsars. PSR~J1703--4851 is shown to be one of a small subclass of pulsars that has a weak and a strong mode which alternate rapidly in time. PSR~J1114--6100 has the fourth highest |RM| of any known pulsar despite its location far from the Galactic Centre. PSR~J1825--1446 shows variations in both DM and RM likely due to its motion behind a foreground supernova remnant.
\end{abstract}



\section{Introduction}
Results from the long-term monitoring of millisecond pulsars have the potential to do fundamental science, such as tests of theories of gravity \citep{ksm+06,agh+18}, determining the equation of state of matter \citep{afw+13}, and the detection of gravitational waves \citep{srl+15,aab+19}. Such experiments have limitations based on our incomplete knowledge of the pulsar emission mechanism, the pulsar spin-down and on the Galactic magneto-ionised medium through which the radio waves propagate. Therefore, observations over decades of the bulk of the population of radio pulsars are warranted as they allow science to be gleaned both from the behaviour of the pulsar itself and from the properties of the interstellar medium (ISM). As examples of the former, carried out with the Parkes telescope, \citet{ymh+13} compiled glitch statistics, \citet{psj+19,pjs+20} investigated the timing noise properties of pulsars over more than a decade and \citet{jk18} investigated the polarization properties of 600 pulsars. Meanwhile, changes in pulse profiles with time \citep{lhk+10} and the implications thereof were discussed in \citet{bkj+16} and \citet{khjs16}. The properties of the interstellar medium were explored through long-term changes in the dispersion measure \citep{pkj+13}, extreme scattering events \citep{kcw+18} and variations in flux density \citep{hdjk+20} for a large sample of pulsars.

The Parkes radio telescope has been observing and timing pulsars over the past 30 years at a variety of observing frequencies and employing signal processing techniques with increasingly better capabilities. In the 1990s and early 2000s, analogue filterbanks operating at an observing frequency of 1400~MHz were used with timing programs described in e.g. \citet{jml+95} and \citet{wmp+00}. Subsequently, digital filterbanks with better frequency resolution and full polarization capability were routinely employed. In 2007, a major timing programme of young, energetic pulsars was set up \citep{wjm+10} to provide ephemerides for NASA's \textit{Fermi} Gamma-ray Space Telescope mission
 \citep{sgc+08}. The programme used a central observing frequency of 1369~MHz and a bandwidth of 256~MHz to monitor a sample of some 150 pulsars on a monthly basis with twice-yearly observations at 3.1~GHz and 0.7~GHz. The programme achieved its goal of increasing the sample of $\gamma$-ray pulsars more than ten-fold \citep{waa+10,2pc, sbc+19}. In late 2018, the Ultra-Wideband receiver (UWL) was installed on the Parkes telescope. The UWL allows observing over the entire band between 704 and 4032~MHz. The FPGA-based digital filterbanks were replaced by a GPU-based software system known as Medusa which enables fully flexible backend configurations to be used. A comprehensive description of the capabilities of the UWL and Medusa can be found in \citet{uwl20}.
\begin{figure*}
\begin{center}
\begin{tabular}{cccc}
\includegraphics[width=4cm,angle=0]{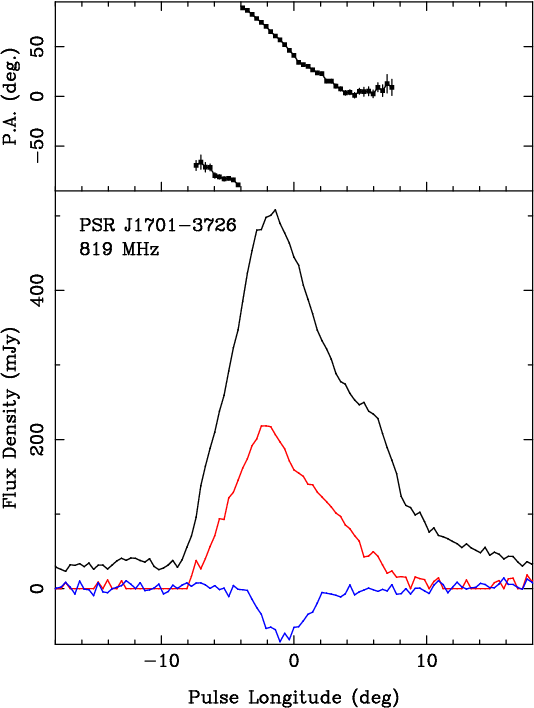} &
\includegraphics[width=4cm,angle=0]{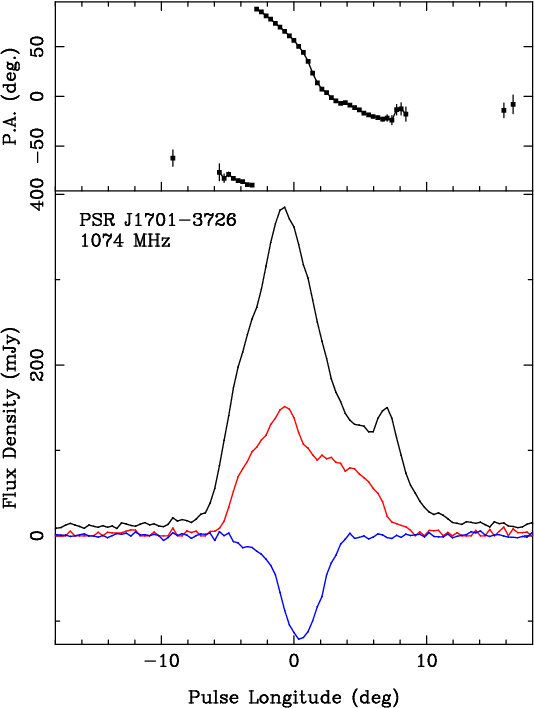} &
\includegraphics[width=4cm,angle=0]{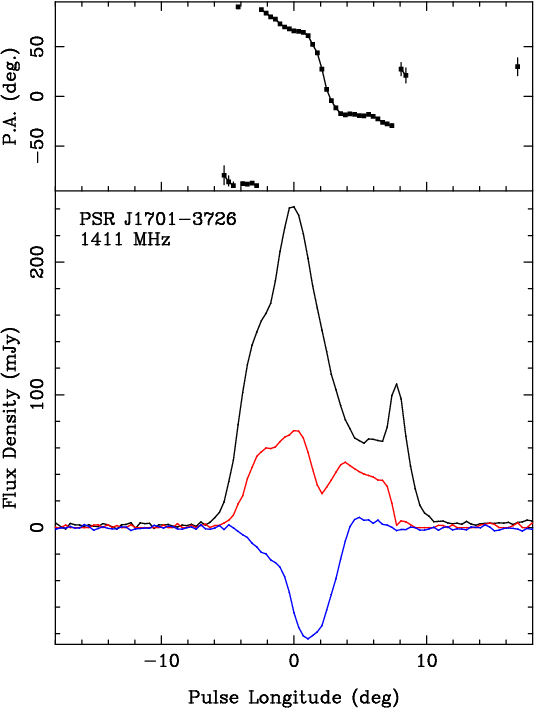} &
\includegraphics[width=4cm,angle=0]{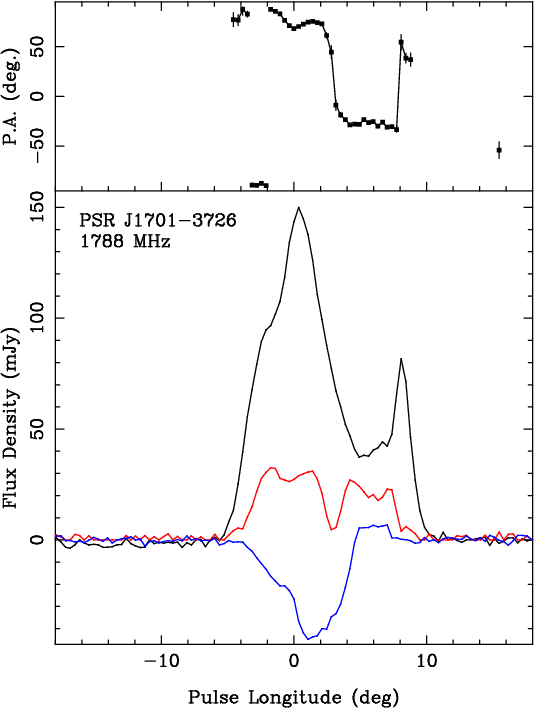} \\
\includegraphics[width=4cm,angle=0]{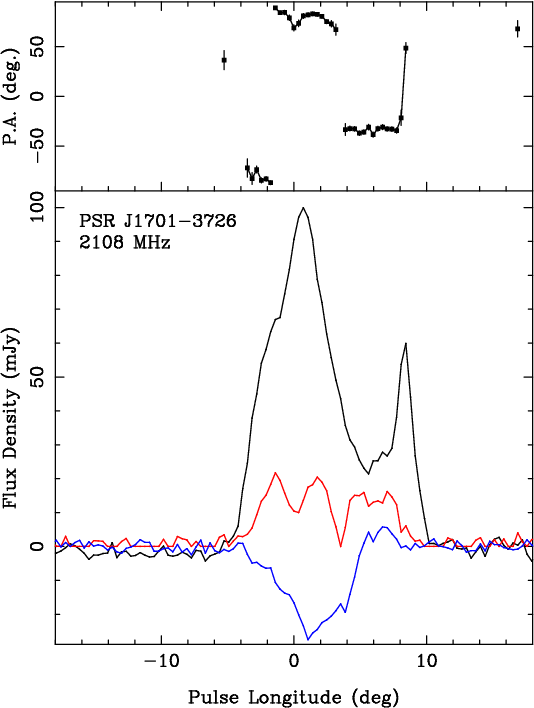} &
\includegraphics[width=4cm,angle=0]{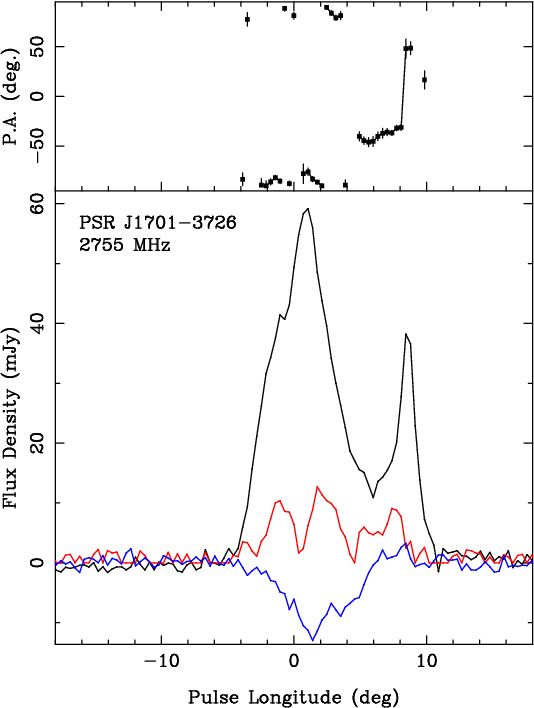} &
\includegraphics[width=4cm,angle=0]{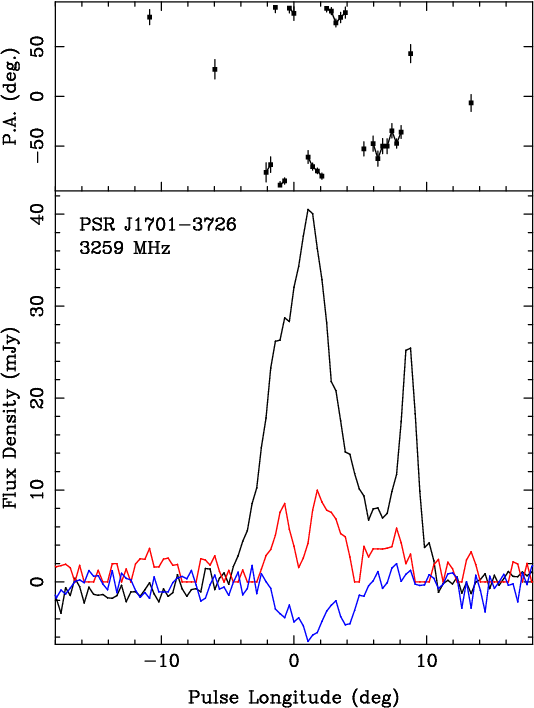} &
\includegraphics[width=4cm,angle=0]{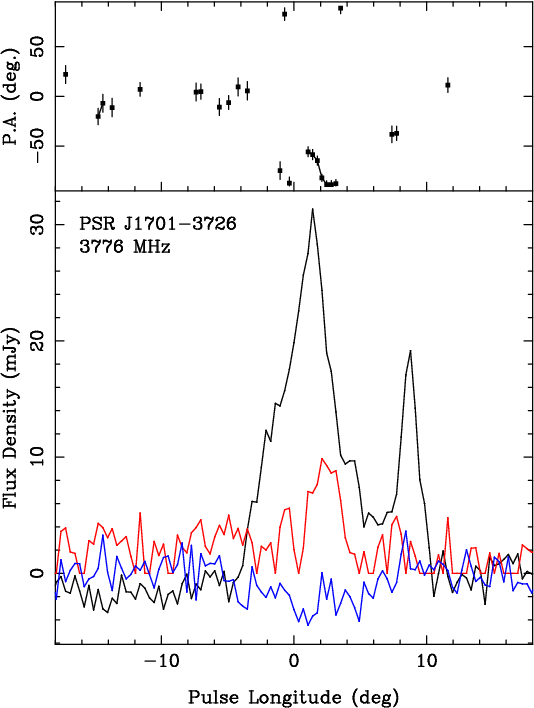} \\
\end{tabular}
\end{center}
\caption{Polarization profiles for PSR~J1701--3726 at eight different frequencies across the UWL band. Black line denotes total intensity, red and blue lines are the linear polarization and circular polarization. The position angle of the linear polarization as a function of pulse phase is also shown.}
\label{fig:1701}
\end{figure*}

In this paper we present results from the first 24 months of young pulsar observations with the UWL, mainly concentrating on aspects of the time-variability of the sample. A polarization study of the brightest pulsars is presented in a companion paper \citep{sjd+21}. Section~2 describes the observations. In Section~3 we look at profile variations with time, section~4 looks at flux density variability, with sections 5 and 6 examining dispersion measure (DM) and rotation measure (RM) changes respectively.

\section{Observations and data reduction}
The initial impetus for this observation programme was the launch of the \textit{Fermi} satellite in 2008. The programme started observing some 150 pulsars with high spin-down energy loss rates, $\dot{E}$, derived from the lists given in \citet{sgc+08}. In 2014 the programme changed emphasis. Observations of some of the weaker, high $\dot{E}$ pulsars were discontinued as it was evident that they were not $\gamma$-ray emitters. A
substantial number of bright, lower $\dot{E}$ pulsars were added when it was realised that these too could be $\gamma$-ray pulsars \citep{sbc+19}. Since 2014 therefore, 276 pulsars are observed in each monthly session in a single block of duration 21~hours. The list of pulsars is given in Table~\ref{tab:applist}.

Observations with the UWL commenced in 2018 November and have occurred roughly monthly thereafter for a total of 23 sessions up to 2020 October. The band between 704 and 4032~MHz is subdivided into 3328 frequency channels each with a channel bandwidth of 1~MHz. Data for each channel are coherently dedispersed at the dispersion measure (DM) of the pulsar and folded at the topocentric spin period to form a pulse profile with 1024 bins across the pulsar period. Data are integrated over 30~seconds and written to disk. The typical observation length for each pulsar is 180~s with a handful of weaker pulsars observed for up to 480~s. Every 60~minutes, observations are made of a pulsed, square-wave, calibration signal to allow for polarization calibration. Flux calibration is carried out via observations of the source Hydra~A. A complete end-to-end description of the system can be found in \citet{uwl20}.

The data are all processed using {\sc psrchive} \citep{hwm04} and data reduction proceeds as follows. First, the calibration observations are examined and radio frequency interference (RFI) signals are removed. The pulsar observation is then calibrated. The square-wave calibration signal provides polarization calibration through correction of the gains and phases. The observation of the flux calibrator converts the digitiser units to mJy. Finally the technique described in \citet{wvs13} is used to account for the instrumental leakage terms. A two-step process is then used to remove RFI from the pulsar data. First the data are summed in time and the RFI flagged in the frequency channels using a median-filter technique and then the individual time steps are examined and bad time integrations are flagged before once again the data are summed in time.

As an example of the output of the data processing, the polarization profiles of PSR~J1701--3726 at eight different frequencies across the band of the UWL are shown in Figure~\ref{fig:1701}.

\section{Pulse profile variability}
\begin{figure}
\begin{center}
\begin{tabular}{c}
\includegraphics[width=7cm,angle=0]{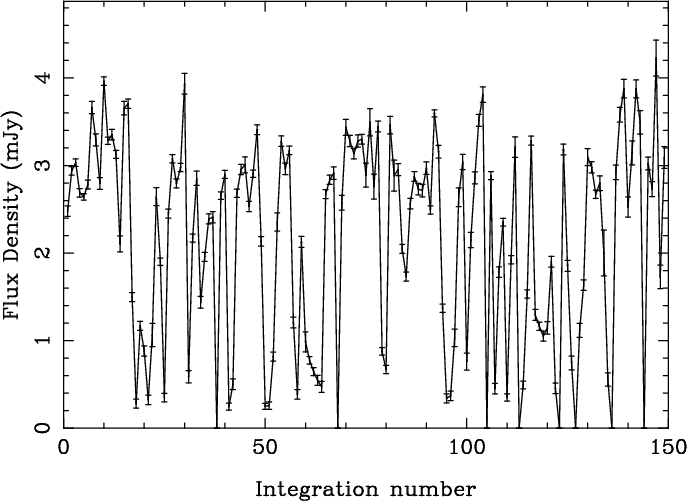} \\
\end{tabular}
\end{center}
\caption{The flux density of PSR~J1701--3726 in each 30~s integration. The integrations dominated by nulling are clearly seen.}
\label{fig:1701b}
\end{figure}
\begin{figure*}
\begin{center}
\begin{tabular}{cc}
\includegraphics[width=8cm,angle=0]{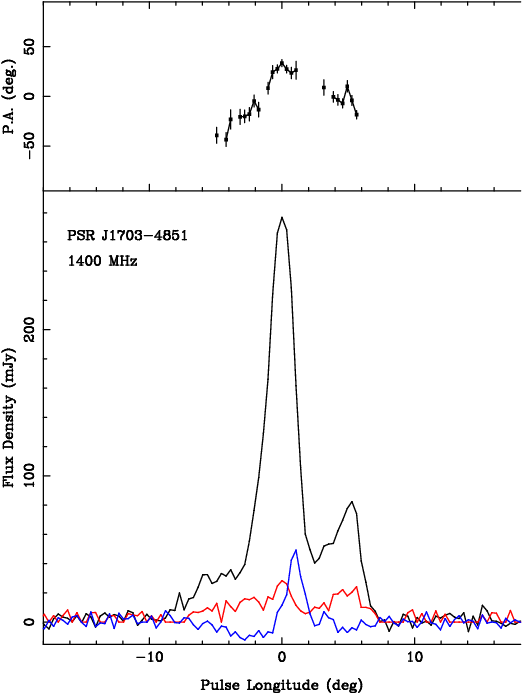} &
\includegraphics[width=7.35cm,angle=0]{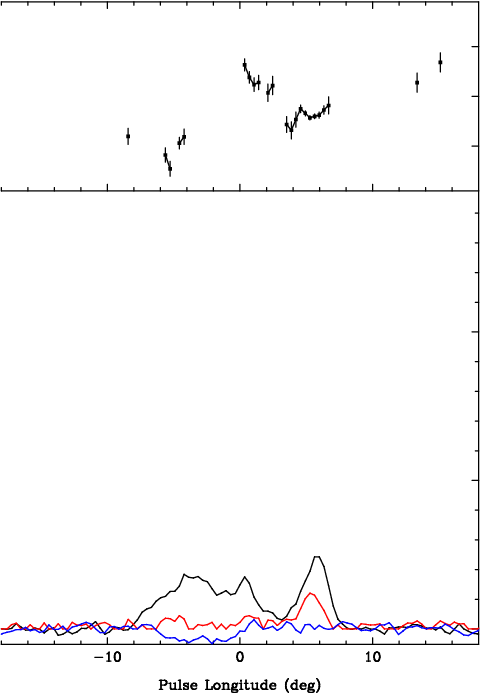} \\
\end{tabular}
\end{center}
\caption{The two modes of PSR~J1703--4851. Left panel shows the bright mode with a dominant central component. The right panel shows the weak mode. The central observing frequency is 1400~MHz with a total bandwidth of 256~MHz.}
\label{fig:1703}
\end{figure*}
For an individual pulsar, each rotation provides a snap-shot pulse profile and these can differ significantly from pulse to pulse. Once sufficient pulses are averaged, a characteristic pulse profile emerges which (at least to zeroth order) appears to be stable over decades. It is this stability which makes high-precision timing experiments possible. However, even in the early days of pulsar research, a phenomenon loosely known as `mode changing' was observed in some pulsars \citep{bac70b}. In these cases, a given pulsar appears to have two or more different, stable pulse profiles. Often, one of the modes is the off or nulling state (e.g. \citealt{bac70,big92}), while in some cases a pulsar may have a `bright' mode and a `quiet' mode (e.g. \citealt{hhk+13}) and in yet other cases have modes of different widths \citep{lhk+10,bkj+16}. The timescale for switching between modes can be as short as one rotation period \citep{syh+15} and as long as years \citep{klo+06}, but what sets the timescale has not yet been determined. Furthermore, without sufficient sensitivity to single pulses, it is unclear whether all pulsars exhibit mode-changing or whether this is restricted to a particular class of pulsar. If we ignore the effects of mode-changing, then pulsar profiles tend to stabilise to within a few percent after several hundred rotations, with a small minority taking much longer to reach a stable profile.

In this particular set of observations we are limited by the fact that we produce profiles every 30~s and so short-term mode-changing is difficult to detect. We also only observe once per month, making it difficult to ascertain the timescale for mode changing. In spite of these caveats, we can measure the flux density for every 30~s integration and inspect the time series for variability. We describe below those pulsars which show nulling and/or other time variable behaviour.

\noindent
{\bf PSR~J0729--1836}: If sufficient rotations are obtained over a long integration time, the pulse profile has a strong leading component followed by a plateau before the trailing drop-off. There is little evolution with frequency. The profiles obtained in individual 180~s observations (350 rotations) can look very different due to the strong and independent modulation of the leading and trailing components. 

\noindent
{\bf PSR~J0820--4114}: This pulsar has a low $\dot{E}$ of $5.4\times10^{30}$~erg\,s$^{-1}$ and a very wide profile which spans more than 100\degr\ of longitude, and it therefore appears to be an old, almost aligned rotator. It often appears to null for about 30~s and remain on for the rest of the 3~min observation so its nulling fraction is less than 15\% which is low compared to other aligned rotators.

\noindent
{\bf PSR~J1048--5832}: The profile of this pulsar has at least four components \citep{sjd+21} and even though we average together 1450 rotations per 3-min observation, the profile can look very different from one observation to the next. We surmise that each component has very different amplitude statistics with occasional bright pulses so that different components dominate depending on the observation length. Single pulse studies are warranted.

\noindent
{\bf PSR~J1049--5833}: This is a long-period (2.2~s) pulsar with a simple, gaussian-like profile. Its nulls typically last for one third of the observation.

\noindent
{\bf PSR~J1114--6100}: This pulsar shows short-duration nulls, with a null fraction less than 20\%.

\noindent
{\bf PSR~J1428--5530}: The profile of the pulsar has two components which are blended and the profile does not stabilise within the 180~s observation either in total intensity or in the fraction of linear polarization. The pulsar sometimes nulls within the 30~s sub-integration time and has a low nulling fraction overall.

\noindent
{\bf PSR~J1646--6831}: This pulsar has as spin period of 1.8-s and a low $\dot{E}$ of $1.2\times10^{31}$erg\,s$^{-1}$. The profile appears to show a classic double-conal structure although the sweep of PA is highly distorted. The circular polarization exceeds the linear polarization fraction in the centre of the profile. The pulsar has short-duration nulls of which several can be seen per observation.

\noindent
{\bf PSR~J1701--3726}: This is a long period (2.5~s) pulsar with a complex profile which varies with frequency (see Figure~\ref{fig:1701}). It nulls within the 30~s integration and is in the null state approximately 25\% of the time. As a result of the nulling, the profile takes a long time to stabilise. Figure~\ref{fig:1701b} shows the flux density in each 30~s integration where the nulling is clearly seen.

\noindent
{\bf PSR~J1703--4851}: This pulsar was flagged in \citet{hdjk+20} as showing unusual behaviour in its flux density variations which were clearly intrinsic to the pulsar rather than a propagation effect. In our sample, it also stands out as an exception. The pulsar shows two distinct modes. The left panel of Figure~\ref{fig:1703} shows the profile of the pulsar in the `bright' mode with a strong central component and weak flanking components. The linear polarization is low and the circular polarization changes sign under the main component. In the `weak' mode (right panel of Figure~\ref{fig:1703}) the central component is reduced in flux density by about a factor of 10 compared to the bright mode and the flanking components stay roughly constant. Of the 14 observations we have made of this pulsar, five are in the bright mode and nine in the weak mode. In one observation, made on 2019 July 6, the pulsar changes from the bright mode to the weak mode in the middle of the observation indicating a fast switching time and a relatively short duration for each mode. This pulsar merits closer attention, a single pulse analysis would prove beneficial.

\noindent
{\bf PSR~J1709--1640}: The low DM of this pulsar means that it exhibits diffractive scintillation, but the timescale of the scintillation is longer than the observation duration. The pulsar has short duration nulls, but is occasionally in the null state during the entire 3~min observation. We estimate the nulling fraction to be 10\%. Both \citet{njmk18} and \citet{whh+20} obtain similar results, but also see occasional nulls with a duration of several hours.

\noindent
{\bf PSR~J1727--2739}: This pulsar nulls with a typical duration of 30~s or longer, and remains on for typically twice this duration. The overall nulling fraction is therefore about 35\%.

\noindent
{\bf PSR~J1733--3716}: This pulsar has a peculiar profile with two separated components each with the steep edge facing the middle of the profile. The leading component is brighter and modulates strongly, both components switch off simultaneously and the nulls are quasi-periodic and of short duration. A single pulse study of this pulsar is given in \citet{hly+20}.

\noindent
{\bf PSR~J1745--3040}: This pulsar also has a peculiar profile which consists of a small leading component almost disjoint from a complex trailing component. Although nulling is not evident, various parts of the profile seem to switch off at various times and the flux in the leading component is often anti-correlated with the flux of the trailing components.

\noindent
{\bf PSRs~J0034--0721, J1825--0935 and J1830--1059} are also part of our sample. These pulsars have a significant literature and their details will not be repeated here (see e.g. \citealt{iwjc20,gjk+94,slk+19} respectively).

\section{Flux densities and spectra}
\begin{figure}
\begin{center}
\includegraphics[width=8cm,angle=0]{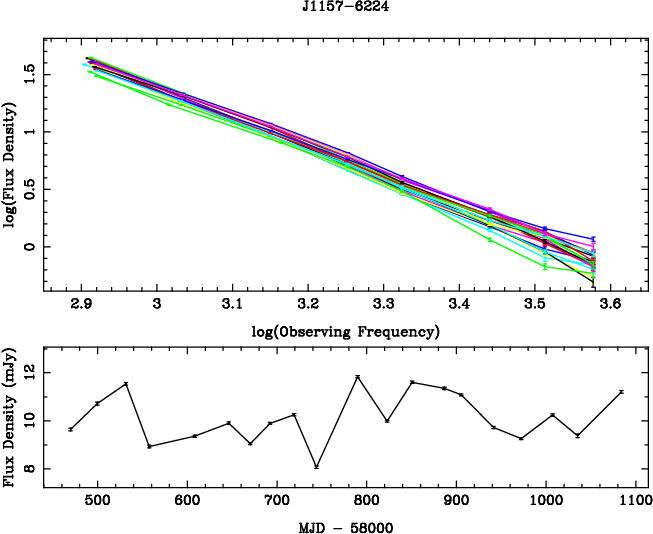}
\end{center}
\caption{Top panel: Flux density versus observing frequency on a log-log scale for PSR~J1157--6224 over 20 epochs. The spectral index is $-2.6.\pm0.05$. Bottom panel: Flux density at 1.4~GHz as a function of epoch.}
\label{fig:J1157flux}
\end{figure}
The UWL allows us to measure the flux density and the spectrum of a pulsar simultaneously over a wide bandwidth and to repeat this on a monthly basis. Flux densities are measured by supplying a template to the {\sc psrchive} routine {\sc psrflux}. Figure~\ref{fig:J1157flux} shows the example of PSR~J1157--6224, measured over 20 epochs. The spectral index is $-2.6\pm0.05$, fluctuations in the flux density are likely the result of refractive scintillation \citep{hdjk+20} in this high DM (325~cm$^{-3}$\,pc) pulsar.

A comprehensive survey of pulsar flux densities at a wide range of frequencies was carried out by \citet{jvk+18}. In that paper they identified five different spectral types (i) simple power-law (ii) broken power-law (iii) log-parabolic (iv) power-law with high-frequency cutoff and (v) power-law with low-frequency turn-over. Many of the pulsars in our sample were classified in that paper and the results will not be repeated here. However, our sample contains 44 pulsars not in the \citet{jvk+18} paper. For these pulsars, we summed together all the observations to smooth over the effects of diffractive and refractive scintillation. We then measured the flux density of each pulsar at 1400~MHz and examined their spectra. The spectrum for both a simple power-law and a log-parabolic can be described by
\begin{equation}
    {\rm log_{10}} S_{\nu} = a [{\rm log_{10}} (\nu/\nu_{0})]^2 + b [{\rm log_{10}} (\nu/\nu_{0})] + c
\end{equation}
where $S_{\nu}$ is the flux density at a frequency $\nu$ and $\nu_{0}$ is 1400~MHz. For the power-law case, $a=0$ and $b$ yields the spectral index. For the log parabolic case, $a$ is the curvature and $b$ gives the local value of the spectral index at $\nu_0$. Table~\ref{tab:spectra} gives the results.

Of the 44 pulsars, only two show evidence for a low-frequency turnover in the spectrum. PSRs~J0631+1036 and J1751--3323 have spectra which peak near 1~GHz, but the limited data below this frequency do not allow us to fully quantify the spectrum. We therefore simply quote the spectral index for the power-law spectrum above 1~GHz in Table~\ref{tab:spectra}. Of the rest, 7 are best fit with a log-parabolic spectrum and 35 with a power-law. The fraction of pulsars in the different classes is in line with the results found by \citet{jvk+18}.

\begin{table}
\caption{Flux density and spectral information for 44 pulsars. Columns 2 and 3 show the flux density at 1400~MHz ($S_{\rm 1400}$) and the associated uncertainty, $\sigma_S$. Column 8 denotes whether the fit is a power-law (PL) or log-parabolic (LP). In the former case, $a$ denotes the spectral index. For the latter case, $a$ is the curvature parameter and $b$ the spectral index, with $\sigma_a$ and $\sigma_b$ their respective uncertainties. PL-T denotes a low frequency turnover, see text for details.}
\label{tab:spectra}
\begin{center}
\begin{tabular}{crrllrrc}
\hline
\hline
Jname & $S_{\rm 1400}$ & $\sigma_S$ & $a$ & $\sigma_a$ & $b$ & $\sigma_b$ & Type\\
 & (mJy) & (mJy) \\
\hline
J0525+1115 & 1.94 & 0.02 & --2.27 & 0.08 & . & . & PL \\
J0631+1036 & 1.11 & 0.01 & --0.33 & 0.03 & . & . & PL-T \\
J0738--4042 & 112.58 & 0.54 & --1.02 & 0.10 & --1.38 & 0.03 & LP \\
J0842--4851 & 1.07 & 0.01 & --1.78 & 0.06 & . & . & PL \\
J1012--5857 & 1.91 & 0.01 & --1.26 & 0.08 & --1.24 & 0.02 & LP \\
J1017--5621 & 2.20 & 0.01 & --0.93 & 0.16 & --2.13 & 0.05 & LP \\
J1046--5813 & 1.37 & 0.01 & --2.19 & 0.10 & . & . & PL \\
J1110--5637 & 3.29 & 0.02 & --0.85 & 0.11 & --1.23 & 0.03 & LP \\
J1114--6100 & 5.36 & 0.02 & --0.7 & 0.1 & --0.53 & 0.03 & LP \\
J1210--5559 & 1.27 & 0.01 & --2.17 & 0.03 & . & . & PL \\
J1225--6408 & 1.26 & 0.01 & --2.12 & 0.12 & . & . & PL \\
J1306--6617 & 4.91 & 0.02 & --1.6 & 0.2 & --1.25 & 0.05 & LP \\
J1418--3921 & 0.95 & 0.01 & --2.52 & 0.04 & . & . & PL \\
J1428--5530 & 7.67 & 0.01 & --2.3 & 0.1 & . & . & PL \\
J1430--6623 & 13.60 & 0.02 & --1.8 & 0.1 & . & . & PL \\
J1544--5308 & 5.82 & 0.01 & --1.71 & 0.06 & . & . & PL \\
J1555--3134 & 4.24 & 0.03 & --0.60 & 0.15 & . & . & PL \\
J1557--4258 & 3.14 & 0.01 & --2.56 & 0.06 & . & . & PL \\
J1559--4438 & 41.06 & 0.05 & --2.3 & 0.1 & . & . & PL \\
J1602--5100 & 8.23 & 0.03 & --2.08 & 0.03 & . & . & PL \\
J1603--5657 & 0.93 & 0.01 & --2.06 & 0.04 & . & . & PL \\
J1623--4256 & 2.60 & 0.02 & --2.35 & 0.05 & . & . & PL \\
J1633--4453 & 2.76 & 0.01 & --2.2 & 0.1 & . & . & PL \\
J1645--0317 & 25.76 & 0.03 & --2.6 & 0.2 & . & . & PL \\
J1649--3805 & 1.75 & 0.02 & --1.75 & 0.08 & . & . & PL \\
J1651--5255 & 2.72 & 0.02 & --2.14 & 0.06 & . & . & PL \\
J1652--2404 & 1.44 & 0.01 & --2.06 & 0.05 & . & . & PL \\
J1705--1906 & 5.66 & 0.03 & --1.54 & 0.04 & . & . & PL \\
J1716--4005 & 1.79 & 0.02 & --1.52 & 0.06 & . & . & PL \\
J1720--2933 & 1.69 & 0.01 & --2.05 & 0.07 & . & . & PL \\
J1722--3632 & 2.90 & 0.02 & --0.7 & 0.3 & --1.1 & 0.1 & LP \\
J1733--2228 & 4.21 & 0.02 & --2.7 & 0.1 & . & . & PL \\
J1735--0724 & 2.76 & 0.01 & --2.28 & 0.02 & . & . & PL \\
J1739--3131 & 7.04 & 0.04 & --2.20 & 0.03 & . & . & PL \\
J1750--3157 & 1.40 & 0.02 & --1.68 & 0.04 & . & . & PL \\
J1751--3323 & 1.67 & 0.02 & --0.92 & 0.03 & . & . & PL-T \\
J1816--2650 & 2.38 & 0.02 & --2.70 & 0.02 & . & . & PL \\
J1817--3837 & 2.09 & 0.01 & --1.82 & 0.06 & . & . & PL \\
J1820--0427 & 10.07 & 0.02 & --2.45 & 0.01 & . & . & PL \\
J1825--0935 & 11.86 & 0.07 & --1.5 & 0.1 & . & . & PL \\
J1834--0426 & 19.69 & 0.08 & --1.60 & 0.02 & . & . & PL \\
J1841--0345 & 2.07 & 0.07 & --1.0 & 0.1 & . & . & PL \\
J1845--0434 & 2.92 & 0.02 & --0.9 & 0.1 & . & . & PL \\
J2155--3118 & 0.66 & 0.02 & --3.0 & 0.1 & . & . & PL \\
\hline
\end{tabular}
\end{center}
\end{table}

\section{Dispersion measure and variability}
\begin{figure}
\begin{center}
\includegraphics[width=8cm,angle=0]{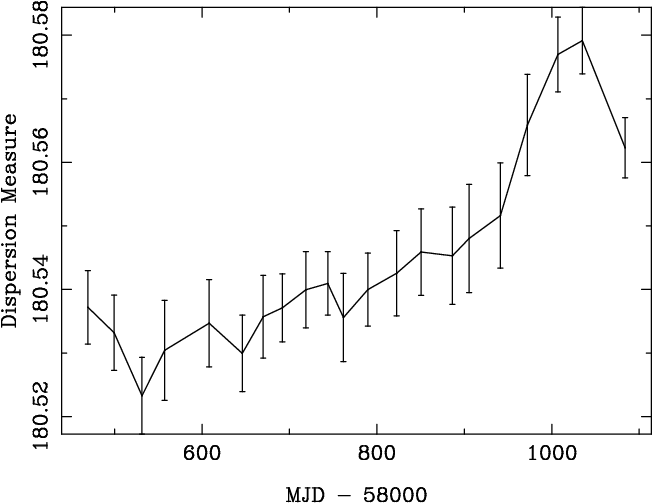}
\end{center}
\caption{Dispersion measure versus MJD for PSR~J0908--4913. A straight line fit to the data yields a slope of $0.026\pm0.02$~cm$^{-3}$\,pc\,yr$^{-1}$.}
\label{fig:J0908dm}
\end{figure}
In this section, we concentrate on changes in DM over time, a quantity which is relatively simple to measure compared to the absolute DM value. Absolute DMs are difficult to measure because of profile variations with frequency and our lack of understanding of magnetospheric processes (see e.g. \citealt{okj20}). For millisecond pulsars, high precision to DM variations can be obtained \citep{pen19}. \citet{yhc+07} and \citet{kcs+13} took advantage of this to examine the DM structure functions for a sample of 20 millisecond pulsars. The previous large-scale study of DM variability in slow pulsars was carried out by \citet{pkj+13} who analysed 168 pulsars over an almost 6-year period. They were able to detect significant variability in only four objects and placed upper limits on the rest. The work of \citet{pkj+13} involved a mixture of narrow-band observations and non-simultaneous observations over several bands and had significant overlap with the pulsars under consideration here.

In order to measure a DM per epoch, we average over the observation duration, and reduce the number of frequency channels to eight across the band. We then use {\sc psrchive} to produce arrival times (ToAs) via the routine {\sc pat} and fit the DM to these ToAs using {\sc tempo2}. We discard observations which are strongly affected by RFI. We then fitted a straight line to the DM as a function of time. Currently 56 of the pulsars in our sample have an upper limit on DM changes of 0.01~cm$^{-3}$\,pc\,yr$^{-1}$, already an improvement on \citet{pkj+13}, and which bodes well for the future of this project as the time baseline increases. We find 11 pulsars with slopes greater than 5-$\sigma$ deviant from zero as listed in Table~\ref{tab:dmdot}. Three of these objects (PSRs~J0835--4510, J0908--4913 and J1833--0827) also had significant values of $\Delta$DM in \citet{pkj+13} but these were of opposite sign compared to our observations. This demonstrates that a simple slope is a crude tool to measure DM variations and that either a higher-order polynomial or a structure function analysis is required (e.g. \citealt{dvt+20}).

\begin{table}
\caption{Pulsars with significant values of $\Delta$DM. The uncertainty in the slope is given in brackets and refers to the last digit(s).}
\label{tab:dmdot}
\begin{center}
\begin{tabular}{crr}
\hline
\hline
Jname & DM & $\Delta$DM \\
& (cm$^{-3}$\,pc) & (cm$^{-3}$\,pc\,yr$^{-1}$) \\
\hline
J0835--4510 & 67.8 & --0.024(2)\\
J0908--4913 & 180.5 & +0.026(2)\\
J1048--5832 & 128.7 & --0.054(4)\\
J1105--6107 & 271.4 & +0.063(4)\\
J1602--5100 & 170.8 & --0.019(3)\\
J1611--5209 & 127.3 & --0.020(3)\\
J1825--1446 & 352.7 & +0.169(20)\\
J1826--1334 & 231.5 & +0.084(11)\\
J1832--0827 & 301.1 & +0.045(6)\\
J1833--0827 & 410.4 & +0.284(23)\\
J1835--1106 & 132.6 & +0.027(4)\\
\hline
\end{tabular}
\end{center}
\end{table}

\section{Rotation measure and variability}
We note that, as with DM, values of rotation measure (RM) for a given pulsar depend on the method used to obtain them, and can typically vary by several units. In addition, an incorrect DM can also lead to an erroneous RM \citep{ijw19}. In order to determine the RM we employ two different algorithms as implemented within the {\sc psrchive} routine {\sc rmfit}. The first method uses a `brute-force' approach in which trial RMs are employed and the amount of linear polarization is computed for each trial value. The algorithm returns the RM at which the linear polarization is maximised over the profile as a whole. In the second method, rotation measures are computed using the algorithm outlined in \citet{njkk08}. In brief, a single position angle (PA) is computed per frequency channel after collapsing the pulse profile to a single phase bin. The rotation measure (RM) is then computed through a quadratic fit to the PA as a function of frequency and the statistical error determined through Monte Carlo trials.

There are nine pulsars in our sample which have had RMs measured for the first time. These are listed in Table~\ref{tab:nocat}. PSR~J1643--4505, once the high RM is taken into account, joins the ranks of pulsars with high $\dot{E}$ and very high linear polarization fraction. In contrast, the linear polarization fraction for the high $\dot{E}$ pulsar PSR~J1055--6028 remains low even after the RM correction. We note the extremely high value of RM uncovered for PSR~J1114--6100, described in more detail in the subsection below. We have found a substantial number of pulsars for which our value of RM is more than 10~rad\,m$^{-2}$ away from the published values as given in v1.63 of the ATNF pulsar catalogue \citep{mht+05}\footnote{http://www.atnf.csiro.au/research/pulsar/psrcat/} . These pulsars are listed in Table~\ref{tab:badcat}. The published value of PSR~J1413--6141 was seriously in error, applying the correct value reveals that this pulsar has only a moderate linear polarization fraction. For PSR~J0857--4424, the published value from 20 years ago of $-75$~rad\,m$^{-2}$ is erroneous; we measure +162~rad\,m$^{-2}$, a difference of more than 200 units.

We observe that there are sometimes significant differences between the RMs returned by the two methods as can be seen in Figure~\ref{fig:J1048rm}. \citet{kar09} pointed out that the effects of interstellar scattering can lead to erroneous measurements of RM, and that the method of \citet{njkk08} was best to use in the case of scattered profiles. In addition, \citet{ijw19} showed the presence of (apparent) RM variations across the pulse profile which meant that different RMs can be obtained depending on the method used. The most extreme example of this is PSR~J1600--5044 for which we obtain an RM of 90~rad\,m$^{-2}$ and 135~rad\,m$^{-2}$ using the two methods. At low frequencies the pulsar is heavily scattered, at high frequencies the position angle of the linear polarization has a very steep swing across the pulse profile. This is the scenario pointed out by \citet{kar09} as being the most detrimental to measuring correct values of RM. 

\begin{table}
\caption{Pulsars without previous RM measurements. The uncertainty is given in brackets and refers to the last digit.}
\label{tab:nocat}
\begin{center}
\begin{tabular}{cr}
\hline
\hline
Jname & RM  \\
& (rad\,m$^{-2}$)  \\
\hline
J0820--3826 & +122(6) \\
J1055--6028 & +343(2) \\
J1114--6100 & --6729(2)\\
J1515--5720 & +41(7)\\
J1637--4642 & --43.2(5)\\
J1643--4505 & --858(1) \\
J1650--4921 & --208(1)\\
J1716--4005 & --526(3) \\
J1843--0702 & +188(3)\\
\hline
\end{tabular}
\end{center}
\end{table}

\begin{table}
\caption{Pulsars with derived RMs which differ by more than 10~rad\,m$^{-2}$ from the published values, RM$_{\rm cat}$. The uncertainties, $\sigma$ and $\sigma_{\rm cat}$ are also listed.}
\label{tab:badcat}
\begin{center}
\begin{tabular}{crrrrr}
\hline
\hline
Jname & RM & $\sigma$ & RM$_{\rm cat}$ & $\sigma_{\rm cat}$ & ref\\
& (rad\,m$^{-2}$) & (rad\,m$^{-2}$) & (rad\,m$^{-2}$) & (rad\,m$^{-2}$)  \\
\hline
J0855--4644 & +225 & 3 &	+249 & 22 & (3) \\
J0857--4424 & +165 & 4 &	--75 & 20 & (5) \\
J0901--4624 & +267 & 2 &	+289 & 22 & (3) \\
J0954--5430 & +86 & 2 &	+65 & 10 & (3) \\
J1003--4747 & +52 & 2 &	+18 & 4 & (3) \\
J1012--5857 & +39 & 4 &	+74 & 6 & (3) \\
J1015--5719 & +110 & 2 &	+96 & 2 & (6) \\
J1016--5857 & --520 & 2 &	--540 & 3 & (6) \\
J1019--5749 & --334 & 3 &	--366 & 10 & (3) \\
J1034--3224 & --43 & 2 &	--8 & 1 & (5) \\
J1038--5831 & --31 & 1 &	--15 & 10 & (8) \\
J1043--6116 & +189 & 2 &	+257 & 23 & (3) \\
J1049--5833 & +344 & 4 &	+359 & 11 & (3) \\
J1146--6030 & +7 & 2 &	--5 & 4 & (8) \\
J1305--6203 & --478 & 2 &	--436 & 15 & (3) \\
J1327--6222 & --327 & 2 &	--306 & 8 & (3) \\
J1341--6220 & --947 & 1 &	--921 & 3 & (6) \\
J1356--5521 & +89 & 2 &	+101 & 4 & (8) \\
J1410--6132 & +2270 & 5 &	+2400 & 30 & (9) \\
J1412--6145 & --52 & 2 &	--130 & 13 & (6) \\
J1413--6141 & --354 & 4 &	--35 & 10 & (3) \\
J1424--5822 & --643 & 3 &	--625 & 19 & (3) \\
J1534--5334 & +25 & 1 &	--46 & 17 & (3) \\
J1536--5433 & --138 & 3 &	--155 & 13 & (3) \\
J1543--5459 & +167 & 1 &	+28 & 23 & (3) \\
J1544--5308 & --42 & 1 &	--29 & 7 & (5) \\
J1548--5607 & +22 & 1 &	+37 & 10 & (3) \\
J1600--5044 & +134 & 1 &	+119 & 10 & (11) \\
J1602--5100 & +84 & 1 &	+71.5 & 1.1 & (11) \\
J1604--4909 & +7 & 2 &	+34 & 1 & (7) \\
J1611--5209 & --101 & 1 &	--79 & 5 & (8) \\
J1638--4417 & +139 & 3 &	+160 & 25 & (12) \\
J1638--4608 & +363 & 3 &	+335 & 12 & (3) \\
J1640--4715 & --422 & 2 &--411 & 12 & (8) \\
J1648--4611 & --650 & 3 &	--682 & 26 & (3) \\
J1715--3903 & +228 & 1 &	+250 & 15 & (3) \\
J1719--4006 & --204 & 2 &	--218 & 17 & (8) \\
J1720--2933 & +10 & 4 &	+21 & 5 & (2) \\
J1722--3632 & --332 & 2 &	--307 & 8 & (3) \\
J1731--4744 & --446 & 2 &	--429.1 & 0.5 & (11) \\
J1738--3211 & --18 & 2 &	+7 & 9 & (3) \\
J1739--2903 & --301 & 2 &	--236 & 18 & (3)\\
J1739--3023 & --120 & 1 &	--74 & 18 & (3) \\
J1740--3015 & --155.3 & 0.4 &	--168.0 & 0.7 & (8) \\
J1750--3157 & +71 & 1 &	+111 & 8 & (3) \\
J1757--2421 & --27 & 1 &	+16 & 5 & (8) \\
J1801--2304 & --1124 & 2 &	--1156 & 19 & (1) \\
J1806--2125 & 724 & 5 &	+796 & 15 & (4) \\
J1822--4209 & +54 & 10 &	--13 & 9 & (3) \\
J1832--0827 & +12 & 1 &	+39 & 7 & (10) \\
J1833--0827 & --284 & 3 &	--470 & 7 & (10) \\
J1848--0123 & +514 & 3 &	+580 & 30 & (11) \\
J1853--0004 & +627 & 4 &	+648.7 & 4.7 & (4) \\
\hline
\end{tabular}
\end{center}
References: (1) \citet{cmk01}, (2) \citet{hl87}, (3) \citet{hml+06}, (4) \citet{hmvd18}, (5) \citet{hmq99}, (6) \citet{jw06}, (7) \citet{jkk+07}, (8) \citet{njkk08}, (9) \citet{ojk+08}, (10) \citet{rl94}, (11) \citet{tml93}, (12) \citet{wj08b}
\end{table}

\subsection{PSR~J1114--6100}
\begin{figure*}
\begin{center}
\includegraphics[width=12cm,angle=-90]{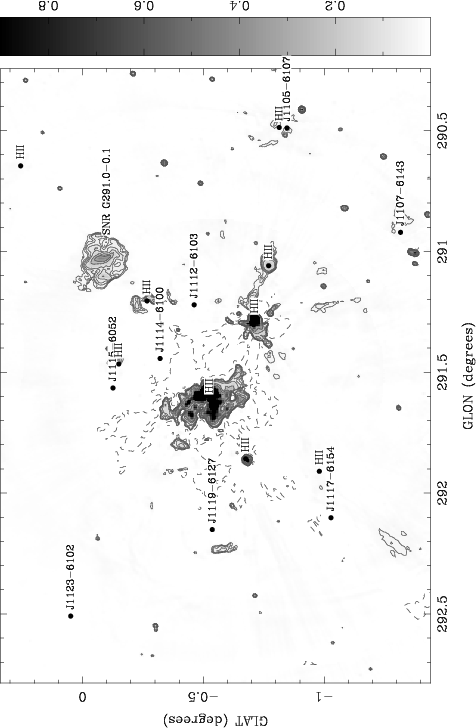}
\end{center}
\caption{Radio continuum image at 843~MHz from the Molonglo Galactic Plane Survey \citep{gcl99} of a region near PSR~J1114--6100. Pulsars and HII regions are marked as is the SNR~G291.0--0.1. The wedge to the right of the image shows the flux density levels in Jy.}
\label{fig:1114}
\end{figure*}
We measured a value of $-6729 \pm 2$~rad\,m$^{-2}$ for the RM of PSR~J1114--6100. This is the fourth highest value of |RM| for any known pulsar, beaten only by the Galactic centre magnetar PSR~J1745--2900 \citep{sj13}, and two other pulsars near the Galactic centre, PSRs~J1746--2849 and J1746--2856 \citep{sef+16}. The DM of the pulsar is 677~cm$^{-3}$\,pc, and so the value of the magnetic field strength parallel to the line of sight, given by 1.2~RM/DM, is $-12.0~\mu$G, the third highest for any pulsar. 

Figure~\ref{fig:1114} shows the location of PSR~J1114--6100 and other pulsars overlaid with a radio continuum image taken from the Molonglo Galactic Plane Survey \citep{gcl99}. Of the pulsars close on the sky to PSR~J1114--6100, most have positive values of RM, including a value of $+853$~rad\,m$^{-2}$ for PSR~J1119--6127. PSRs~J1112--6103 and J1115--6052 both have RMs close to $+250$~rad\,m$^{-2}$. The prominent supernova remnant SNR~G291.0--0.1 \citep{rjgk06} lies close on the sky to the pulsar and there are numerous HII regions in the vicinity \citep{ch87}. The largest and brightest HII region located at (l,b = 291.614,--0.525) has a flux density in excess of 200~Jy; its recombination lines have a positive velocity indicating a distance of some 8~kpc. This HII region (also known as NGC~3603) is seen from radio to gamma-rays \citep{mcs+02,sdt+20}, and is the subject of a significant body of literature; a recent study of its OB stars places its distance at 7~kpc \citep{dmw19}. To the south-west lies the only HII region (291.284,--0.713) with negative velocities at a distance of some 4~kpc. In the optical, the HII regions are clearly delineated and do not extend as far as PSR~J1114--6100. H$\alpha$ images show some excess diffuse structure near the pulsar. The pulsar shows a relatively low amount of scatter-broadening given the large value of DM.

The cause of the extremely high RM of PSR~J1114--6100 is unclear. Its distance, derived from the DM \citet{ymw17}, is roughly consistent with the distance to NGC~3603 but the HII region appears to be too distant (in angle) to contribute to the DM of pulsar. It could be that the pulsar lies behind some highly magnetised filament located perhaps in the near part of the Carina spiral arm. In any case, it remains odd that the pulsar's RM is an order of magnitude greater than its neighbouring pulsars, and pulsars in other, more complex regions of the Galactic plane.

\subsection{Time-variable RMs}
\begin{figure}
\begin{center}
\includegraphics[width=8cm,angle=0]{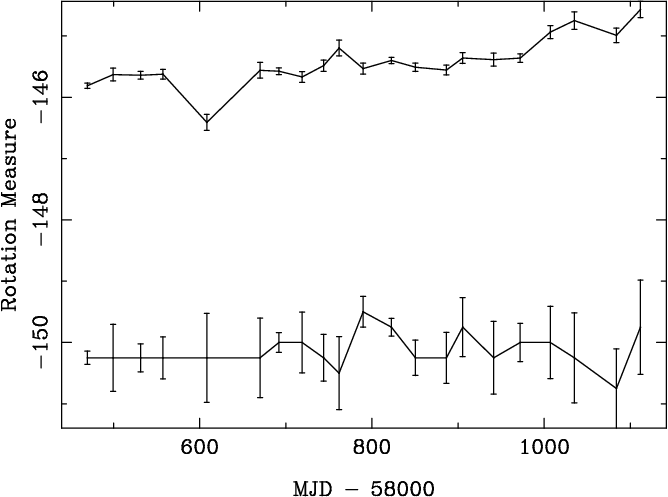}
\end{center}
\caption{Rotation measure versus epoch for PSR~J1048--5832. Top curve shows the RM using the `brute-force' method, bottom curve shows the RM using the Noutsos et al. method.}
\label{fig:J1048rm}
\end{figure}
There are 10 pulsars in our sample for which the statistical error in the measured RM is smaller than 1.0~rad\,m$^{-2}$ per epoch. These are the brightest and/or the most highly polarized pulsars, PSRs~J0738--4042, J0742--2822, J0835--4510, J0908--4913, J1048--5832, J1359--6038, J1644--4559, J1709--4429, J1740--3015 and J1745--3040. In no case is there evidence for any change in RM. As an example, Figure~\ref{fig:J1048rm} shows RM versus epoch for PSR~J1048--5832. The error bars per epoch reflect statistical errors and do not take into account the ionospheric RM. However, there does appear to be one pulsar which does show RM changes, PSR~J1825--1446, a pulsar which we noted earlier also showed DM changes. This pulsar is discussed further below. The pulsar with the second largest $|$RM$|$ in our sample is PSR~J1410--6132, which has a value of +2400~rad\,m$^{-2}$ according to \citet{ojk+08} but for which we measured +2279~rad\,m$^{-2}$, a substantial change. Our data show a decline in RM from 2290~rad\,m$^{-2}$ to 2270~rad\,m$^{-2}$ over the course of 2 years although the significance of the change is only 2-$\sigma$. However, it could be the case that the RM has decreased substantially since 2008 and further monitoring is required.

\subsection{PSR~J1825--1446}
\begin{figure}
\begin{center}
\begin{tabular}{c}
\includegraphics[width=8cm,angle=0]{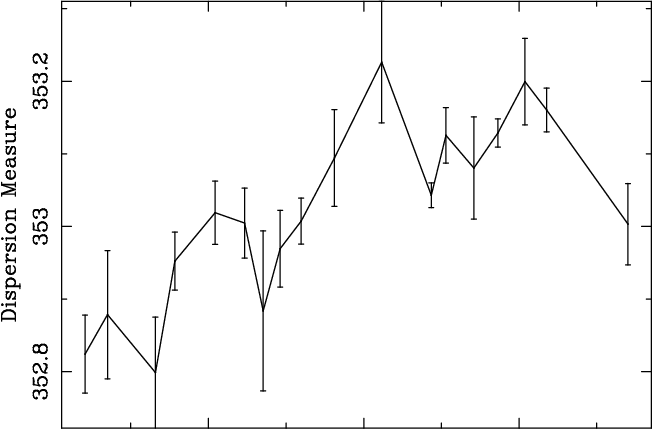} \\
\includegraphics[width=8cm,angle=0]{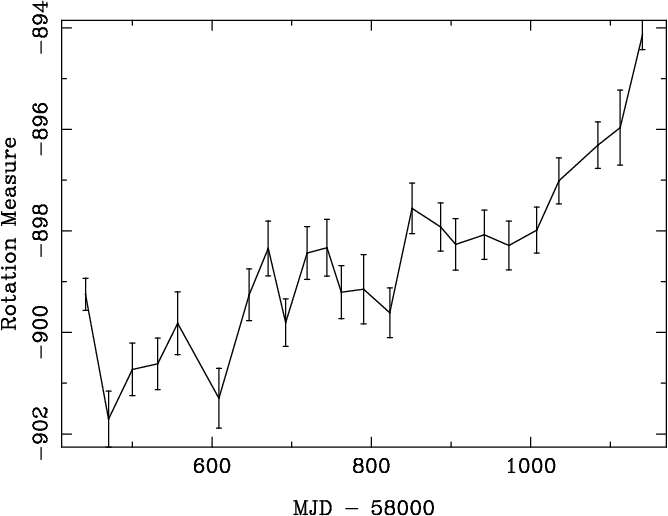} \\
\end{tabular}
\end{center}
\caption{Dispersion measure (top panel) and rotation measure (bottom panel) as a function of epoch for PSR~J1825--1446. A straight line fit to the DM versus time yields a slope of $0.15\pm0.02$~cm$^{-3}$\,pc\,yr$^{-1}$, that to the RM versus time yields $2.6\pm0.3$~rad\,m$^{-2}$\,yr$^{-1}$.}
\label{fig:1825}
\end{figure}

This pulsar is the only one of our sample which shows significant changes in both RM and DM, shown in Figure~\ref{fig:1825}. The RM increases almost linearly by about 8~rad\,m$^{-2}$ over two years (much larger than expected from the ionosphere) while the DM increases by about 0.3~cm$^{-3}$\,pc. The implies a change in the magnetic field along the line of sight of 0.2~$\mu$G in two years.
An independent distance to this pulsar has not been determined but the distance inferred from the DM is $\sim$5~kpc. The proper motion is high, as measured by \citet{mrp+12} and later refined by \citet{ddb+17} and hence the pulsar has a high transverse velocity of 750~kms$^{-1}$. The pulsar also appears in projection to be located within the shell of the supernova remnant G16.8--1.1 although \citet{mrp+12} argue that the pulsar's age and velocity make a true association unlikely. Nevertheless, our line of sight to the (background) pulsar intercepts the SNR and it seems likely that the RM and DM changes we see are the result of the pulsar passing behind a magnetised filament in the SNR. The scattering time-scale of the pulsar is 21~ms at 1~GHz \citep{osw21}. This is an order of magnitude higher than expected from the extrapolation of the relationship of \citet{kmnj+15} to 1~GHz, but is not particularly discrepant compared to other pulsars with comparable DMs \citep{osw21}.

\section{Summary}
The Ultra-Wideband receiver on the Parkes telescope has been operational for two years and will remain the workhorse for pulsar observations through this decade. A complementary programme of slow pulsar observations on the MeerKAT telescope is also underway \citep{mtime,tpa20} and will be key for single-pulse analysis. Repeated monitoring of a large sample of pulsars coupled with the wide instantaneous observing bandwidth is ideal for studying the time-evolution of the pulsars themselves and the interstellar medium through which their radio emission propagates. The early results presented here are extremely encouraging for the future of this project. We have reported on a number of new nulling pulsars, DM and RM variability, and a pulsar with an extremely high value of RM.

\section*{Acknowledgements}
The Parkes radio telescope is part of the Australia Telescope National Facility which is funded by the Australian Government for operation as a National Facility managed by CSIRO. RMS acknowledges support through Australian Research Council Future Fellowship  FT190100155. Work at NRL is supported by NASA. We thank the referee Scott Ransom for his heartening report.

\section*{Data Availability}
Pulsar data taken for the P574 project is made available through the CSIRO's Data Access Portal (\href{https://data.csiro.au}{https://data.csiro.au}) after an 18 month proprietary period.



\bibliographystyle{mnras}
\bibliography{uwl1} 


\appendix
\section{Pulsar list}
\begin{table*}
\caption{List of the 276 pulsars monitored with the UWL on a monthly basis.}
\label{tab:applist}
\begin{center}
\begin{tabular}{cccccccccc}
\hline
\hline
PSR \\
\hline
J0034--0721 & J0108--1431 & J0134--2937 & J0151--0635 & J0152--1637 & J0206--4028 & J0255--5304 & J0304+1932  & J0401--7608 & J0448--2749 \\
J0452--1759 & J0525+1115  & J0536--7543 & J0543+2329  & J0601--0527 & J0614+2229  & J0624--0424 & J0627+0706  & J0630--2834 & J0631+1036 \\
J0659+1414  & J0729--1448 & J0729--1836 & J0738--4042 & J0742--2822 & J0745--5353 & J0758--1528 & J0809--4753 & J0820--1350 & J0820--3826 \\
J0820--4114 & J0834--4159 & J0835--4510 & J0837+0610  & J0837--4135 & J0842--4851 & J0855--4644 & J0857--4424 & J0901--4624 & J0904--7459 \\
J0905--5127 & J0907--5157 & J0908--1739 & J0908--4913 & J0924--5814 & J0940--5428 & J0942--5552 & J0954--5430 & J0959--4809 & J1001--5507 \\
J1003--4747 & J1012--5857 & J1015--5719 & J1016--5819 & J1016--5857 & J1017--5621 & J1019--5749 & J1028--5819 & J1034--3224 & J1038--5831 \\
J1043--6116 & J1046--5813 & J1047--6709 & J1048--5832 & J1049--5833 & J1055--6028 & J1056--6258 & J1057--5226 & J1105--6107 & J1110--5637 \\
J1112--6103 & J1114--6100 & J1115--6052 & J1119--6127 & J1123--6259 & J1136--5525 & J1146--6030 & J1156--5707 & J1157--6224 & J1210--5559 \\
J1224--6407 & J1225--6408 & J1243--6423 & J1253--5820 & J1301--6305 & J1302--6350 & J1305--6203 & J1306--6617 & J1317--6302 & J1319--6056 \\
J1320--5359 & J1326--5859 & J1326--6408 & J1326--6700 & J1327--6222 & J1327--6301 & J1328--4357 & J1338--6204 & J1340--6456 & J1341--6220 \\
J1349--6130 & J1352--6803 & J1356--5521 & J1357--62   & J1357--6429 & J1359--6038 & J1401--6357 & J1410--6132 & J1412--6145 & J1413--6141 \\
J1418--3921 & J1420--6048 & J1424--5822 & J1428--5530 & J1430--6623 & J1435--5954 & J1452--6036 & J1453--6413 & J1456--6843 & J1509--5850 \\
J1512--5759 & J1513--5908 & J1515--5720 & J1522--5829 & J1524--5625 & J1524--5706 & J1530--5327 & J1531--5610 & J1534--5334 & J1534--5405 \\
J1535--4114 & J1536--5433 & J1539--5626 & J1541--5535 & J1543--5459 & J1544--5308 & J1548--5607 & J1549--4848 & J1555--3134 & J1557--4258 \\
J1559--4438 & J1600--5044 & J1600--5751 & J1602--5100 & J1603--5657 & J1604--4909 & J1605--5257 & J1611--5209 & J1613--4714 & J1614--5048 \\
J1623--4256 & J1626--4537 & J1630--4733 & J1632--4621 & J1633--4453 & J1633--5015 & J1637--4553 & J1637--4642 & J1638--4417 & J1638--4608 \\
J1638--4725 & J1640--4715 & J1643--4505 & J1644--4559 & J1645--0317 & J1646--4346 & J1646--6831 & J1648--3256 & J1648--4611 & J1649--3805 \\
J1649--4653 & J1650--4502 & J1650--4921 & J1651--4246 & J1651--5222 & J1651--5255 & J1652--2404 & J1653--3838 & J1653--4249 & J1700--3312 \\
J1701--3726 & J1701--4533 & J1702--4128 & J1702--4310 & J1703--3241 & J1703--4851 & J1705--1906 & J1705--3423 & J1705--3950 & J1707--4053 \\
J1707--4729 & J1708--3426 & J1709--1640 & J1709--4429 & J1715--3903 & J1715--4034 & J1716--4005 & J1717--3425 & J1718--3825 & J1719--4006 \\
J1720--2933 & J1721--3532 & J1722--3207 & J1722--3632 & J1722--3712 & J1723--3659 & J1727--2739 & J1730--3350 & J1731--4744 & J1733--2228 \\
J1733--3716 & J1735--0724 & J1737--3137 & J1738--3211 & J1739--2903 & J1739--3023 & J1739--3131 & J1740--3015 & J1741--2733 & J1741--3016 \\
J1741--3927 & J1743--3150 & J1745--3040 & J1749--3002 & J1750--3157 & J1751--3323 & J1751--4657 & J1752--2806 & J1757--2421 & J1801--2304 \\
J1801--2451 & J1803--2137 & J1806--2125 & J1807--0847 & J1809--1917 & J1816--2650 & J1817--3618 & J1817--3837 & J1820--0427 & J1822--2256 \\
J1822--4209 & J1823--3106 & J1824--1945 & J1825--0935 & J1825--1446 & J1826--1334 & J1828--1101 & J1829--1751 & J1830--1059 & J1832--0827 \\
J1833--0827 & J1834--0426 & J1835--0643 & J1835--1106 & J1837--0559 & J1841--0345 & J1841--0425 & J1842--0905 & J1843--0702 & J1844--0538 \\
J1845--0434 & J1845--0743 & J1847--0402 & J1848--0123 & J1852--0635 & J1852--2610 & J1853--0004 & J1900--2600 & J1910+0358  & J1913--0440 \\
J1932+2220  & J1941--2602 & J2048--1616 & J2155--3118 & J2330--2005 & J2346--0609 \\
\hline
\end{tabular}
\end{center}
\end{table*}

\bsp	
\label{lastpage}
\end{document}